# Growth Process of Hexagonal Boron Nitride in the Diffusion and Precipitation Method Studied by X-ray Photoelectron Spectroscopy

Satoru Suzuki*, and Yuichi Haruyama

*Laboratory of Advanced Science and Technology for Industry, University of Hyogo, Koto, Kamigori, Ako, Hyogo 678-1205, Japan*

E-mail: ssuzuki@lasti.u-hyogo.ac.jp

Submonolayer h-BN was grown on Ni foil in ultra-high vacuum by the diffusion and precipitation method and the growth process was studied by x-ray photoelectron spectroscopy. Formation of h-BN started to be observed at 600 °C. All through the process, the surface was always slightly B-rich, which is consistent with the fact that B which is soluble in Ni at a high temperature can diffuse in Ni by the conventional bulk diffusion and insoluble N cannot. Moreover, both formation and decomposition of h-BN were found to occur at elevated temperatures possibly depending on provision of N atoms to the surface. On the Ni surface, decomposition of h-BN was observed at a relatively low temperature of 800 °C.

## 1. Introduction

Hexagonal boron nitride (h-BN) is very promising as a substrate material for graphene devices[1-5] and as a tunneling barrier layer of tunneling devices.[6-11] In addition to the one-atom thickness and the large band gap (~6 eV), the large chemically stability of h-BN is one of very important properties for various applications. The thermal decomposition temperature of bulk h-BN has been reported to be extremely high (2127-2327 °C).[12] It has also been demonstrated that an atomically thin h-BN film protects a Ni substrate from oxidation in oxygen atmosphere even up to 1100 °C.[13]

Recently, chemical vapor deposition (CVD) is commonly used for growth of large-area and atomically thin h-BN.[14-17] CVD growth is normally performed under gas flow with relatively high or ambient pressure. Therefore, direct characterization of the growth process by electron spectroscopy is very difficult. In fact, there have been few reports on spectroscopic observation of the growth process of h-BN CVD. Kidambi et al., reported in-situ observation of CVD growth by x-ray diffraction and x-ray photoelectron spectroscopy (XPS).[18] They studied growth of multilayer h-BN film on Cu foil.

The diffusion and precipitation method is an alternative way to grow atomically thin h-BN film.[19-22] This method utilizes only solid phase reaction of a metal (Ni was used in this study) substrate and solid source (a-BN). B and N atoms diffuse from the bottom surface to the top surface of the metal foil at an elevated temperature and form h-BN on the top surface. Therefore, h-BN can be grown in an ultra-high vacuum (UHV). Thus, this method is very suitable for analysis of growth process in a UHV environment. In the diffusion and precipitation method, it would be expected that feeding mechanism of B and N atoms to the surface is different. Although B is known to be slightly soluble in Ni at high temperature,[23] the solubility of N is negligibly small even at high temperature.[23] [24-26] Grain boundary diffusion (in Ni) and surface migration (on top surface) were considered to be responsible for feeding N atoms from our previous studies.[20,21]

In this paper, we studied the growth process of monolayer h-BN in the diffusion and precipitation method by using x-ray photoelectron spectroscopy (XPS). Unbalanced feeding of B and N atoms to the surface was directly observed. Moreover, we show that h-BN is decomposed on Ni at a relatively low temperature of 800 °C.

## 2. Experimental methods

A submonolayer h-BN film was grown on Ni foil using the diffusion and precipitation method, based on our previous studies.20,21) A amorphous boron nitride (a-BN) film (thickness: 100 nm) was deposited on a Ni foil (Nilaco., thickness: 25 μm) using radio frequency magnetron sputtering in $Ar/N_2$ atmosphere ($Ar/N_2$ flow rate: 1:1). The a-BN/Ni sample was loaded in the sample preparation chamber connected to the XPS analysis chamber. In the preparation chamber (base pressure: $4\times10^{-8}$ Pa), the sample was heated at a certain temperature in a range of 600 to 1000 °C by electron bombardment to the a-BN-deposited face. The maximum pressure observed during heating did not exceed $4\times10^{-6}$ Pa (at 1000 °C). Then, the sample was transferred to the analysis chamber ($4\times10^{-8}$ Pa) and XPS of undeposited face was measured at room temperature. Unmonochromatized Al Kα line (1486.6 eV) and a hemi-spherical electron energy analyzer (Scienta, R-3000) were used for XPS measurements. Heating and XPS measurements were repeated, as schematically shown in Fig. 1. Heating temperature and time at each process were summarized in Table 1. In this study, the maximum h-BN coverage was observed at process 6 and estimated to be about 0.5 from B 1s (N 1s)/Ni 2p XPS intensity ratio.

## 3. Results and discussion

### 3.1 Results

A wide range XPS of the surface of the Ni foil before heating is shown in Fig. 2. The surface of the Ni foil was oxydized in air. The surface was also contaminated by hydrocarbon as denoted by the C 1s signal. The N 1s signal was observed because the surface was once exposed to nitrogen radicals during the reactive sputtering. Although the observed surface did not face to $Ar/N_2$ plasma, nitrogen radials enveloped the foil. In fact, N 1s signal is not observed when the a-BN deposition is performed in pure Ar atmosphere.[19] These N atoms initially observed before heating are considered to be non-essential for h-BN growth, because h-BN can be grown even if sputter-deposition of a-BN is performed in pure Ar atomosphere.[19-21] In fact, the N atoms were mostly removed after heating at 600 °C for 10 min as also shown in Fig. 2. Almost no O 1s signal is observed after heating meaning that Ni surface was reduced. Simultaneously, hydorocarbon also disappeared. The Ni 2p intensity increased because bare Ni appeared on the surface after heating.

B 1s and N 1s XPS obtained at each process are shown in Fig. 3. The N 1s signal centered at 399 eV observed before heating (process 0) is not due to h-BN formation, but surface nitridation during the a-BN deposition, as mentioned above. In fact, the N 1s binding energy differs from that of h-BN (397.6 eV), and no B 1s signal is observed. The spectra shows that h-BN formation slightly started at 600 °C (process 1). In the B 1s spectrum, h-BN formation is still barely visible. However, in the N 1s spectrum, a small peak is observed at about 397.6 eV, which corresponds to N 1s binding energy of h-BN.[18-20)27)] Instead, the peak initially observed at 399 eV disappeared. After annealing at 800 °C and higher temperatures (process 2-12), h-BN formation is also clearly visible in B 1s spectra. However, the B 1s spectra split into two peaks. The peak at 190.0 and 188.0 eV are assigned to h-BN and elemental B, respectively.[18-20)] The results indicate that feeding of B and N atoms are not balanced, but the surfaces are B-rich. The elemental B component is likely to appear on the surface during cooling, because B is slightly soluble in Ni at a high temperature (0.3 % at 1065 °C).[23)] The B-rich surfaces are consistent with our previous conclusion that h-BN formation is restricted by provision of N atoms to the surface.[19-21)] The elemental B is quickly oxidized when the sample is exposed to air, as shown in our previous report.[27)]

As can be seen in Fig. 3, the B 1s and N 1s intensity evolution is not monotonic. Figure 4 shows B 1s and N 1s intensity evolutions obtained from Fig. 3. The B 1s (B-N) and N 1s intensity evolutions are reasonably similar to each other due to h-BN formation. Interestingly, heating sometimes caused B 1s and N 1s intensity decreases. Heating at constant temperatures of 800 (processes 2, 3), 850 (4-8), 900 °C (9, 10), the B 1s and N 1s intensities increased at first, and then, they declined. In this heating process, further temperature increase merely decreased the coverage of h-BN (11, 12). These results indicate that not only formation of h-BN but also decomposition occurs at these temperatures. The decomposition was even observed at a relatively low temperature of 800 °C (process 3).

## 3.2 Discussion

Recently, we could directly observe B atoms solved in Ni foil after h-BN formation by using B-*K* x-ray emission spectroscopy.[27)] This is consistent with the fact that B is slightly soluble in Ni at a high temperature. Thus, B atoms can reach the top surface by conventional bulk diffusion. On the other hand, our previous study strongly suggest that provision of N atoms is caused by grain boundary diffusion and surface migration.[19-21] Figure 5 shows a schematic model for explaining the XPS intensity evolutions observed in this study. Heating causes recrystallization of Ni. Then, N atoms in the a-BN film can reach the top surface through the grain boundary. N atoms migrate on the top surface and form h-BN by bonding to B atoms. This study shows that formation and decomposition of h-BN are competing at the elevated temperatures. At first, the provision of N atoms is relatively large and formation of h-BN is dominant. However, repeated heating (processes 3, 7, 8, 10-12) results in decreased provision of N atoms. Then, decomposition becomes dominant. We think that h-BN layers formed at the a-BN/Ni interface[19] is highly responsible for the decrease of N atom provision. The h-BN layers would act as a barrier for N atoms in the a-BN film to diffuse into the Ni foil. We did not observe any N 1s signal other than N-B bond in XPS except for process 0. Thus, insoluble N atoms created by decomposition are released to vacuum possibly by forming inert $N_2$ molecules. B atoms created by decomposition can either melt into Ni or remain on the surface.

Decomposition of h-BN was observed at a relatively low temperature of 800 °C. This temperature is much lower than the decomposition temperature of bulk h-BN (2127-2327 °C).[12] A catalytic effect of the Ni surface is also considered to be important for decomposition of h-BN at such a low temperature. Using the diffusion and precipitation method, monolayer or few-layer h-BN can be grown on Ni foil typically at 900-1000 °C.[20,27] However, heating time is also important for thickness control. The thickness turns to decrease due to decomposition if heating time is too long.

## 4. Conclusions

Growth process of h-BN on Ni in the diffusion and precipitation method was studied by x-ray photoelectron spectroscopy with keeping the sample in UHV. Formation of h-BN was observed at 600 °C. All through the process, the surface was always slightly B-rich, which is consistent with the fact that B is soluble in Ni at a high temperature and N is insoluble in Ni. Formation and decomposition of h-BN were found to be competing at the elevated temperatures. On the Ni surface, decomposition of h-BN was observed at 800 °C.

## Acknowledgments

We thank Prof. Yuichi Utsumi, Prof. Akinobu Yamaguchi, and Mr. Masaya Takeuchi for their technical support with the a-BN deposition. This work was supported by JSPS KAKENHI (JP16H03835, JP16H06361).

## Figure Captions

**Fig. 1.** Schematic of h-BN formation in UHV by the diffusion and precipitation method (left) and its XPS analysis (right). The sample was heated by electron bombardment using a hot filament.

**Fig. 2.** Wide range XPS of the sample obtained before heating and after heating at 600 °C for 10 min.

**Fig. 3.** (a) B 1s and (b) N 1s XPS obtained at each process. The numbers in parenthesis correspond to the process numbers in Table 1.

**Fig. 4.** (a) B 1s and (b) N 1s integrated XPS intensity at each process. In (a), B 1s intensities from h-BN and elemental B are shown, respectively. In (b), the initial N intensity observed at process 0 was excluded in order to focus on N 1s intensity from h-BN.

**Fig. 5.** Schematic model for formation and decomposition of h-BN. When provision of N atoms through Ni grain boundary is large, formation of h-BN is dominant (left). When provision of N atoms is small, decomposition becomes dominant (right).

**Table I.** Summary of the sequential heating processes.

| Process no. | Heating temperature and time |
|:-----------:|:----------------------------:|
| 0 | Before heating |
| 1 | 600 °C, 10 min |
| 2 | 800 °C, 10 min |
| 3 | 800 °C, 10 min |
| 4 | 850 °C, 10 min |
| 5 | 850 °C, 10 min |
| 6 | 850 °C, 10 min |
| 7 | 850 °C, 10 min |
| 8 | 850 °C, 10 min |
| 9 | 900 °C, 10 min |
| 10 | 900 °C, 20 min |
| 11 | 950 °C, 10 min |
| 12 | 1000 °C, 10 min |

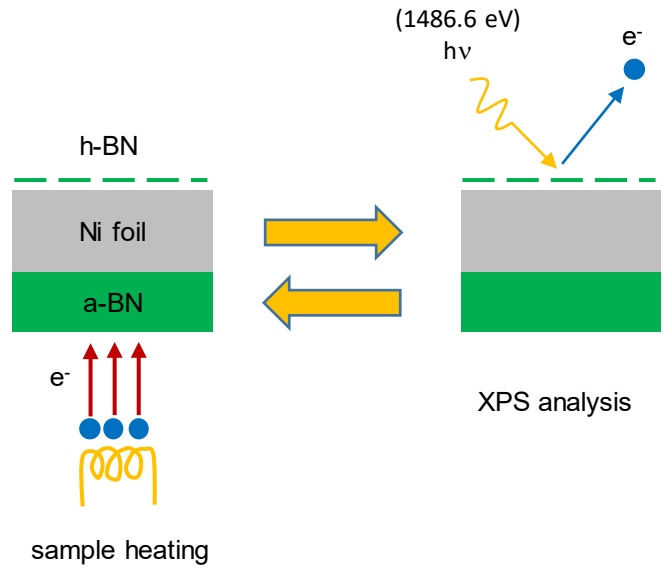

Figure 1

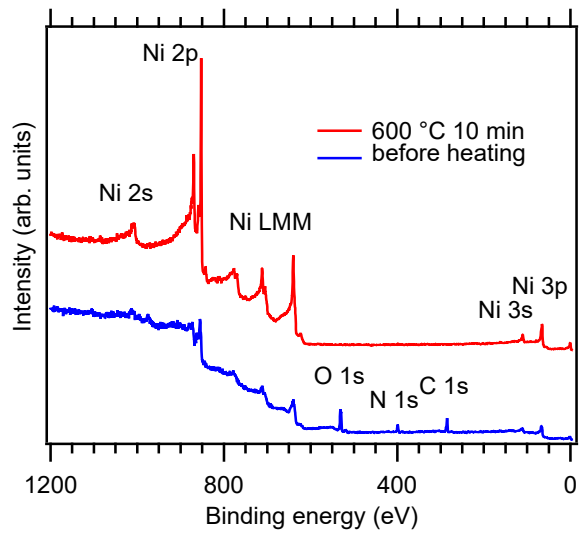

Figure 2

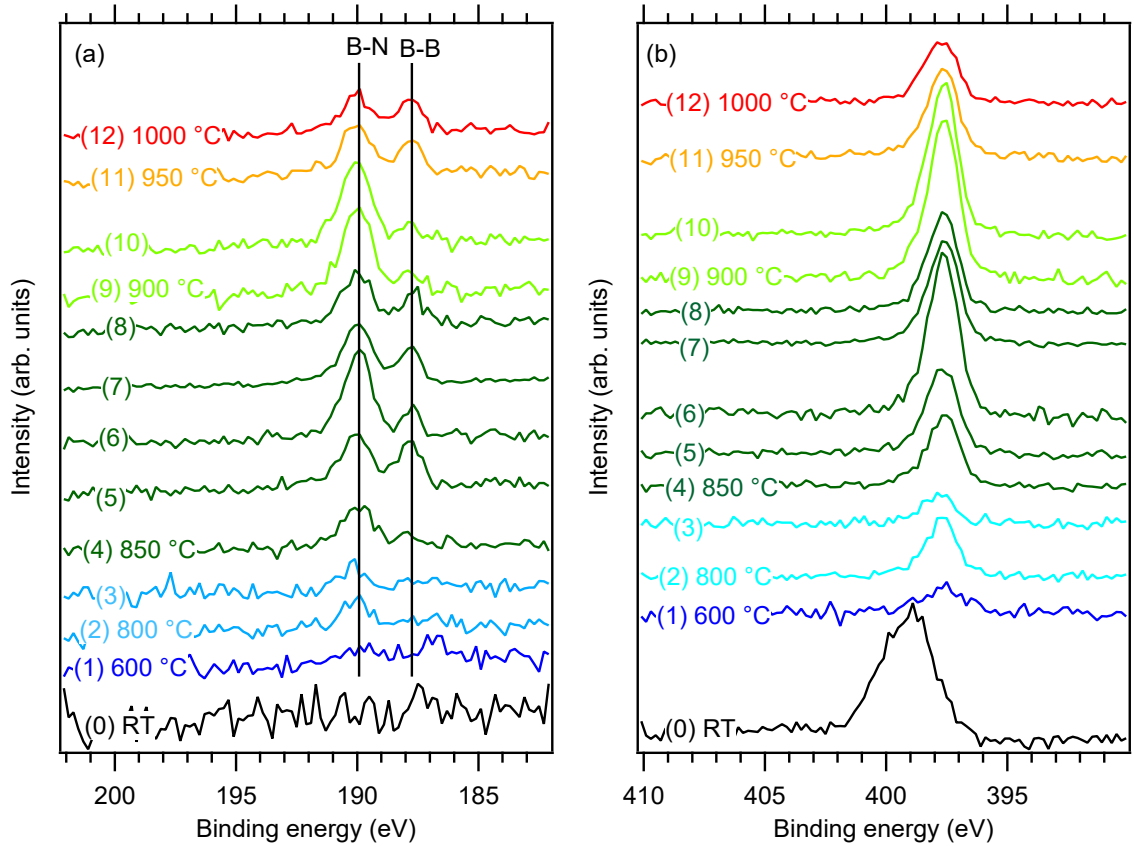

Figure 3

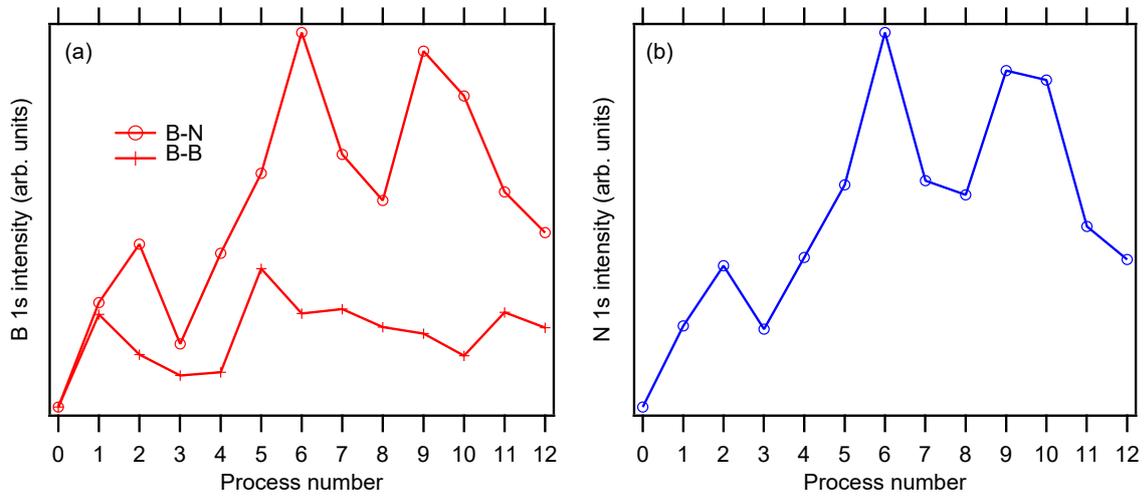

Figure 4

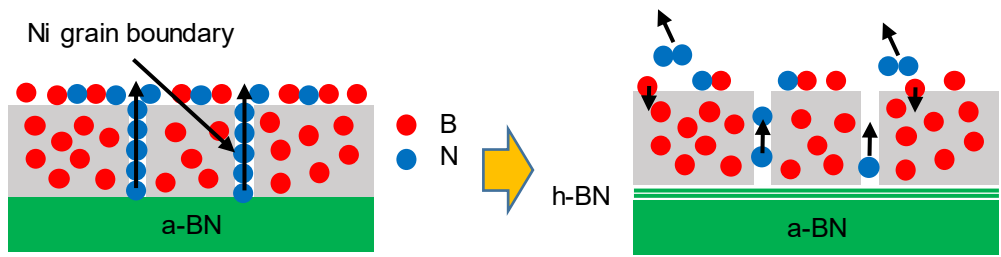

Figure 5